\documentstyle[epsfig,12pt,preprint,tighten,aps]{revtex}
\begin{document}
\draft
\title{
\rightline{{\tt October 2000}}
\
\\ Physics of mirror photons}
\author{R. Foot, A. Yu.\ Ignatiev and R. R. Volkas}
\address{School of Physics\\
Research Centre for High Energy Physics\\
The University of Melbourne\\
Victoria 3010 Australia\\
foot, sasha, r.volkas@physics.unimelb.edu.au }
\maketitle

\begin{abstract}
The physics of kinetic mixing between ordinary and mirror photons
is discussed. An important role is played by four
linear combinations we dub the physical photon, the
sterile photon, the physical mirror photon, and the sterile
mirror photon. Because of the mass degeneracy between the two
gauge bosons, quantum coherence effects are important.
The physical photon becomes a
certain coherent superposition of the bare ordinary photon
and the bare mirror photon. Similarly, the physical mirror
photon is another, but {\it not orthogonal},
coherent superposition. We discuss the physics of the interaction
between physical mirror photons and ordinary matter.
Observational signatures for some hybrid ordinary/mirror binary
astrophysical systems are qualitatively discussed. We show that a small
amount of ordinary matter at the center of a mirror star may make the
mirror star observable.
We speculate that the recently reported halo white
dwarfs might actually be mirror halo stars.
\end{abstract}

\section{Introduction}
\label{intro}

There has been considerable interest in the study of particle
physics implications of the atmospheric \cite{atmos},
solar\cite{solar}, LSND \cite{lsnd}
and other neutrino experiments. The
main purpose of such investigations has been to understand what
theories could be responsible for the observed experimental features,
which in the atmospheric and solar cases
give strong evidence for large angle neutrino oscillations. One
of these theories is the Exact Parity Model (EPM) (also known as
the Mirror Matter Model) which introduces
parity or ``mirror'' partners for all ordinary particles
(except the graviton) and thus
restores the parity invariance apparently broken by the weak
interactions\cite{ly,blin,flv}.
The EPM predicts pairwise maximal mixing between ordinary
and mirror neutrinos and provides a basis for an explanation of
the atmospheric and solar neutrino  data\cite{flv2}.
This explanation does not require large mixing between
generations and is also consistent with the LSND
experiment.\footnote{In the EPM the parity symmetry
is unbroken by the vacuum,
which is one of the reasons it is so simple and predictive.
It is also possible to envisage models where the parity symmetry
is spontaneously broken\cite{spon} which typically have quite
different features.}

Another interesting implication of this idea is that the mirror
nucleons/atoms provide a natural candidate for dark
matter\cite{blin,hod,macho2}.
This hypothesis can nicely explain some of the
features of dark matter. For example, the
heavy MACHO objects inferred to exist
in the halo of our galaxy by the microlensing experiments\cite{macho}
can be naturally interpreted as mirror stars\cite{macho2}.
If the dark matter is composed of mirror matter then it is also
possible that some mirror matter exists in our solar system and
in other solar systems.
In fact a mirror planet could form very close to an ordinary star
and may provide a possible explanation for the close-in extra-solar
planets discovered around nearby stars\cite{plan}. For other possible
astrophysical, cosmological and geophysical implications of mirror matter see
Ref.\cite{cos}.

Ordinary and mirror matter interacts through
gravitation, and through the mixing of
colourless and neutral particles with their mirror
partners\cite{flv,flv2,hol,gl}:
neutrinos, the photon, the Z boson and the neutral Higgs boson can mix
with their corresponding mirror states.\footnote{
Other possibilities require physics beyond the minimal model.
The most interesting of these would appear to be neutron
-- mirror neutron and $K_L - K_L'$
mixing (the former requires baryon number violation, while the latter requires
flavour changing neutral currents which could occur, for example, in a two Higgs
doublet extension).}
Of particular concern to this paper
is photon -- mirror photon kinetic mixing,
\begin{equation}
{\cal L}_{\text{int}} = - \epsilon F_{\mu\nu} F'^{\mu\nu},
\label{Lint}
\end{equation}
where $F_{\mu \nu}$ ($F'^{\mu\nu}$) is the field strength tensor
for ordinary (mirror) electromagnetism, and $\epsilon$ is a
free parameter.
An important experimental consequence
of Eq.(\ref{Lint})
is the mixing of
orthopositronium with mirror orthopositronium\cite{gl},
leading to oscillations between these states in a vacuum experiment.
The subsequent decays of the mirror state
are invisible, resulting in an
effective increase in the decay rate\cite{gl}.
A longstanding discrepancy between the
theoretical prediction and some of the experimental
measurements may in fact be resolved by this mirror
world mechanism\cite{fg} (see also Ref.\cite{gen}).
These experiments suggest the
value $\epsilon \sim 5 \times 10^{-7}$ which is too large
to be palatable for standard BBN\cite{cg}, but is nevertheless very
interesting because of its terrestrially observable consequences.

The purpose of the present work is to consider in detail the
physics of the photon -- mirror photon mixing. The degeneracy
between the ordinary and mirror photons (both of which are assumed to
be massless) leads to coherence effects which should be taken into
account when considering various processes involving ordinary/mirror
photon emission, absorption and scattering.
Various interesting astrophysical implications will also be
qualitatively discussed.
In particular we will find that a small amount of ordinary matter at
the center of a mirror star may make the mirror star observable.
Indeed such objects may already have been observed, but interpreted
as types of white dwarfs.

\section{$U(1) \otimes U(1)'$ quantum electrodynamics}
\label{u1xu1}

Let us consider $U(1)\otimes U(1)'$ quantum electrodynamics containing
an ordinary electron $\psi$ and photon $A$ plus
a mirror electron $\psi'$ and mirror photon $A'$,
\begin{equation}
\label{1}
{\cal L} = -{1 \over 4}F_{\mu\nu}^2-{1 \over 4}F_{\mu\nu}^{'2}
+\bar{\psi}(i\hat{\partial}-m)\psi +\bar{\psi'}(i\hat{\partial}-
m)\psi'
+e\bar{\psi}\hat A\psi + e\bar{\psi'}\hat A'\psi'
-\epsilon F_{\mu\nu}F'^{\mu\nu}.
\end{equation}
Observe that the last term of the Lagrangian describes the kinetic mixing between
the ordinary and mirror photons.
This term is gauge invariant and renormalisable and can exist at tree
level\cite{flv,fh}, or may be induced radiatively in models without
$U(1)$ gauge symmetries (such as grand unified theories)\cite{hol,gl,cf}.
Because of kinetic mixing, it turns out that the physical photon
is a particular linear combination
of $A$ and $A'$ (as will be discussed in the following section)\cite{hol}.
Treating kinetic mixing as a part of the
interaction Lagrangian,
we obtain the usual Feynman rules plus an extra $A - A'$
mixing vertex shown in Fig.\ 1a.

To the lowest order in $\epsilon$, the $A - A'$ propagator
has the form
\begin{equation}
\label{2}
-i{2 \epsilon \over k^2}\left[ g_{\mu \nu} - {k_{\mu}k_{\nu}\over
k^2}\right].
\end{equation}
The interaction between the ordinary electron and $A'$
(also to first order in $\epsilon$) takes the form (Fig.\ 1b):
\begin{equation}
\label{3}
2 \epsilon e \gamma^{\mu}.
\end{equation}
Similarly, the A field interacts with the mirror electron via
the diagram in Fig.\ 1c.

\section{Physical effects of kinetic mixing}
\label{4states}

Clearly one effect of kinetic mixing is to couple mirror
electrons to ordinary electrons with an effective
charge $2\epsilon e$ \cite{hol}.
There are also potentially other effects. For example,
since the $A'$ field interacts very weakly with ordinary
matter one might expect that it could carry
energy away from stellar
interiors thus speeding up their evolution and,
consequently, standard astrophysical arguments would place a
strict upper limit on the coupling constant and therefore the mixing
parameter $\epsilon$\cite{kolb}.

However, one should take
into account the coherence between the emission of the ordinary and
mirror photons $A$ and $A'$. Thus the two diagrams in Figs.\ 1b and 1d should
always be added up coherently. It is convenient to introduce the new
field
\begin{equation}
\label{4}
A_1={A + 2\epsilon A' \over \sqrt{1 + 4\epsilon^2}},
\end{equation}
which can be identified with the {\it physical photon field}.
The orthogonal combination
\begin{equation}
\label{5}
A_2={A' - 2 \epsilon A \over \sqrt{1 + 4\epsilon^2}},
\end{equation}
does not interact with ordinary matter at all.
We call $A_2$ the {\it sterile photon}.

Thus we see that in the interior of ordinary stars only the physical
photon field $A_1$ is emitted or absorbed and the additional degrees
of freedom corresponding to the field $A_2$ are completely decoupled.
This observation has also been made in Ref.\ \cite{hol}.
We therefore conclude that the arguments based on stellar energy loss
cannot give us a constraint on the mixing parameter $\epsilon$, assuming
that the  possible
admixture of mirror matter in the ordinary stars can be ignored.

The same arguments as above applied to the mirror star
instead of the ordinary one show us that the mirror star would emit
the {\it physical mirror photon}
\begin{equation}
\label{6}
A_1^{'}={A' + 2 \epsilon A \over \sqrt{1 + 4 \epsilon^2}},
\end{equation}
whereas the orthogonal field
\begin{equation}
\label{7}
A_2^{'}={A - 2 \epsilon A' \over \sqrt{1 + 4\epsilon^2}},
\end{equation}
will be sterile with respect to mirror matter.
We call $A_2'$ the {\it sterile mirror photon}.
These four states ($A_1, A_2, A'_1, A'_2$) are summarised in Fig.\ 2.

Next, let us discuss how one can detect (in principle)
the radiation emitted by the mirror star.
As explained above, mirror matter will emit the physical
mirror photon state $A'_1$.
To see how the physical mirror photon interacts with
ordinary matter, let us rewrite the field $A'_1$ as a superposition of
the physical photon field $A_1$ and the sterile photon field $A_2$:
\begin{equation}
\label{8}
A_1^{'} = {1 - 4 \epsilon^2 \over 1 + 4 \epsilon^2}A_2 +
{4 \epsilon \over 1 + 4 \epsilon^2}A_1.
\end{equation}
If this superposition between $A_1$ and $A_2$ is maintained, then
the mirror photon field couples to ordinary matter with the
coupling constant
\begin{equation}
\label{9}
g = {4 \epsilon \over  1 + 4 \epsilon^2} e.
\end{equation}
While this superposition is preserved
in vacuum, and in a medium of purely mirror matter,
we will show that in {\it ordinary} matter it may be lost due to
interactions with the medium.

\section{Interaction of physical mirror photons with an ordinary matter medium}
\label{decoherence}

We introduce a $2 \times 2$ reduced density matrix for the neutral
gauge boson system in the $(A_1, A_2)$ basis,
\begin{equation}
\label{10}
\rho = {1 \over 2} (P_0 + \sigma \cdot {\bf P}),
\end{equation}
where ${\bf P}$, according to common usage, is called ``the
polarisation vector'' (note that this vector is completely different
from the ``true'' polarisation vector describing the photon spin state).
If the initial
wave is composed of mirror photons, which are 
described by the pure superposition of Eq.(\ref{8}), then
the corresponding density matrix is
\begin{equation}
\label{11}
\rho(0) =
\left( \begin{array}{cc}
\left( { 4 \epsilon \over 1 + 4\epsilon^2}\right)^2 &
{4 \epsilon (1- 4\epsilon^2) \over (1+ 4\epsilon^2)^2} \\
{4 \epsilon (1- 4\epsilon^2) \over (1+ 4\epsilon^2)^2} &
\left( {1- 4\epsilon^2 \over 1+4\epsilon^2}\right)^2
\end{array}\right),
\end{equation}
where we have taken $t = 0$ and normalised so that $P_0(0)=1$ initially.

Suppose that the initially pure state described by $\rho(0)$ propagates through a
medium of ordinary matter. The physical photon component $A_1$ interacts
with the medium, whereas the $A_2$ component is sterile. This is
similar to an active-sterile two flavour neutrino system propagating
through a plasma \cite{qke}.
The density matrix therefore evolves according to the Bloch-type equation \cite{qke}
\begin{equation}
\label{12}
{d{\bf P} \over dt }= {\bf V}\times {\bf P} - \frac{\Gamma}{2} {\bf P}_T
- \Gamma {1 \over 2}(P_0 + P_z){\hat z},\quad
{dP_0 \over dt}= - \Gamma {1 \over 2}(P_0 + P_z),
\end{equation}
where ${\bf P}_T$ is the projection of the vector ${\bf P}$ on the
$(x,y)$-plane, and the positive quantity
$\Gamma$ is the total interaction rate for
an ordinary photon with the medium (including both elastic and
inelastic processes).
The first term on the right-hand side describes the non-absorptive
evolution of the system. It can be viewed as a precession of the
vector ${\bf P}$ around the direction ${\bf V}$, similar to the
precession of a magnetic moment around the direction of a magnetic
field. The second and third terms on the right-hand side
of Eq.(\ref{12}) describe the effects of interaction with the
medium.

The components $V_x$ and $V_y$ are given by the
off-diagonal elements of the mass matrix and therefore vanish in the basis
$(A_1, A_2)$. So, the Bloch equation simplifies to the coupled system
\begin{eqnarray}
{dP_x \over dt} & = & - V_z P_y - \frac{\Gamma}{2} P_x, \nonumber \\
{dP_y \over dt} & = & V_z P_x - \frac{\Gamma}{2} P_y, \nonumber\\
{dP_z \over dt} & = & - \Gamma {1 \over 2}(P_0 + P_z), \nonumber\\
{dP_0 \over dt} & = & - \Gamma {1 \over 2}(P_0 + P_z),
\label{13}
\end{eqnarray}
where $V_z$ is due to coherent forward scattering (refractive index).

For the simple case of constant $\Gamma, V_z$,
the general solution of Eqs.(\ref{13}) is
\begin{eqnarray}
P_z(t) & = & {1 \over 2}[P_z(0) - 1] + {1 \over 2}[P_{z}(0) + 1] e^{-\Gamma t},
\nonumber \\
P_x(t) & = & e^{- \frac{\Gamma}{2}t}[P_{x}(0)\cos{V_z t} - P_{y}(0)\sin{V_z t}],
\nonumber \\
P_y(t) & = & e^{- \frac{\Gamma}{2}t}[P_{y}(0)\cos{V_z t} + P_{x}(0)\sin{V_z t}],
\nonumber \\
P_0(t) & = & - {1 \over 2}[ P_z(0) - 1] + {1 \over 2} [P_z(0) + 1]e^{-\Gamma t}.
\label{14}
\end{eqnarray}
The effect of interaction with the medium thus forces
the density matrix into the form,
\begin{equation}
\label{15}
\rho=
\left( \begin{array}{cc}
0 & 0 \\
0 & \left( {1- 4\epsilon^2 \over 1 + 4\epsilon^2}\right)^2
\end{array}\right) ,
\end{equation}
exponentially quickly with characteristic timescale $\sim 1/\Gamma$.
This density matrix describes a beam of purely sterile photons $A_2$.

Note, however, that if the medium
is perfectly transparent to ordinary
electromagnetic radiation then there is no collisional/absorptive
interaction with the medium. This means (for example) that in a refracting
telescope, the radiation from a mirror star would behave like ``light''
which refracts very feebly, leading to a very long focal length,
and so great difficulty of detection.

\section{Astrophysical Applications}
\label{astro}

As already discussed in Sec.\ref{4states}, an ordinary star emits just the
physical photon field so there is no observable effect of kinetic
mixing. Further, Secs.\ref{4states} and \ref{decoherence} examined
the case of a mirror star, whose mirror photon flux $F$
will
produce effects equivalent to a flux of approximately $4\epsilon^2 F$ of ordinary
light when encountering non-transparent ordinary matter.
Since $\epsilon \stackrel{<}{\sim} 10^{-6}$\cite{gl,gen} this makes the
mirror star
undetectable with present technology unless there
happens to be a very luminous one very nearby.

However, there are a number of other physical situations which
would be expected to occur
given the diversity and size of the universe.
In particular, let us consider the following four possibilities
\begin{enumerate}
\item Ordinary star with a mirror matter core.
\item Mirror star with an ordinary matter core.
\item Ordinary star with an orbiting mirror planet/star.
\item Mirror star with an orbiting ordinary planet/star.
\end{enumerate}
For the purposes of this paper we will qualitatively
identify some of the implications
of kinetic mixing. A detailed quantitative study is beyond the
scope of the present work, but an interesting topic for the future.

\subsection{Ordinary star with a mirror matter core.}

We assume that the mirror matter is confined within a radius $R' < R$,
where $R$ is the radius of the star.
If $\epsilon$ is nonzero then
the embedded mirror matter will be heated up by the interactions with
the surrounding ordinary photons.
It will lose energy by radiating
mirror photons, which will be able to escape if they are emitted near
the mirror surface at $R'$.
An equilibrium situation will develop with the mirror
matter having some approximate temperature $T'$ at its surface.
As the mirror photons propagate out through the star they will,
through interactions with the medium, very
quickly become sterile photons (with a tiny fraction becoming ordinary
photons). They will then rapidly escape from the star,
thereby cooling it.

\subsection{Mirror star with an ordinary matter core.}

We assume that the ordinary matter is confined within a radius $R < R'$,
where $R'$ is the radius of the mirror star.
This case is just the mirror image of the previous one, so
the dynamics of the system will be the same.
However, in this case, the escaping
radiation consists of sterile mirror photons (rather than the sterile
ordinary photons of the previous case).
The sterile mirror photons will become
ordinary photons when they interact with ordinary matter (with
a tiny fraction becoming sterile ordinary photons) and thus potentially
observable by us here on earth!
Therefore this case is potentially more interesting than the previous
one.
There will be several distinctive signatures of this radiation:
Its luminosity will be small (depending on $R$ and $T$) while
it can potentially be relatively high in frequency because the
surface temperature (i.e.\ at radius $R$) of the ordinary matter, $T$, can
be larger than the surface temperature of the mirror star.
Thus, such an object would have roughly similar phenomenological
characteristics to a type of white dwarf.
In fact, there are types of white dwarf (e.g.\ DC white dwarfs)
whose origin is not well
understood and may be candidates for mirror stars with some
ordinary matter in the centre.
Perhaps a good place to look for mirror stars is in the halo of
our galaxy. Interestingly, several recent papers\cite{ibata} claim to have
discovered faint moving objects in the halo with an
estimated abundance consistent with the MACHO data.
The interpretation of these objects as ordinary white dwarfs leads
to a number of difficulties\cite{diff}. It is therefore tempting
to interpret them as mirror stars with a small component of
ordinary matter in the centre.
This mirror star interpretation can be distinguished
from the conventional white dwarf case if the gravitational
red shift of the spectral lines can be measured.
Light from white dwarfs experience much greater red shift
than the light from the center of a typical star.
Also note that this type of fake white dwarf
could have a mass exceeding the Chandrasekhar limit of 1.4 solar masses:
such an object would be a smoking gun for some sort of invisible
clumped matter that happens to have acquired an ordinary matter core.

\subsection{Ordinary star with an orbiting mirror planet/star}

The ordinary star will emit ordinary photons. Through interactions
with the medium,
these photons will mostly become sterile mirror photons when
they propagate through the mirror planet/star. They will thus
travel right through this object making it
essentially transparent (only a tiny component of order $\epsilon^2$ will
be absorbed). The sterile mirror photons will then become ordinary photons
when they propagate through ordinary matter (which is what
all our detection systems are made of).
However in a realistic system one might expect
the mirror planet/star to have a small amount of ordinary
matter embedded into it
which might have accreted over
time (from e.g. the solar wind).
In this case the mirror planet can potentially be
opaque, because the small amount of ordinary matter can absorb
the ordinary photons from the ordinary star.
In the particular case of the large close-in extra-solar planets,
this may well be expected because they would accrete a significant
amount of ordinary matter from the ordinary star during its lifetime.
A detailed study of the optical properties of such
an object needs to be done in order to understand the opacity and
albedo properties.
Obviously any ordinary matter inside the mirror planet will become
quite hot and could be a detectable source of ordinary radiation.

\subsection{Mirror star with an orbiting ordinary planet}

This system is the mirror image of the previous one.
In this case the light from the mirror star may be undetectable,
but the ordinary light emitted from the orbiting mirror planet
may be observable if such a system is in our neigbourhood
and still young.
Recently, the detection of ``isolated planetary mass objects''
has been claimed\cite{iso}. In a separate work, we have proposed that
these objects might actually not be isolated planets, but rather
ordinary planets orbiting invisible mirror stars\cite{fiv}. This idea can be
tested by looking for a periodic Doppler shift in the radiation
emitted by the planet.

\section{Conclusions}

The physics of kinetic mixing between ordinary and mirror photons
has been discussed. An important role is played by four
linear combinations we have dubbed the physical photon, the
sterile photon, the physical mirror photon, and the sterile
mirror photon. Because of the mass degeneracy between the two
gauge bosons, quantum coherence effects have to carefully
considered. In particular, the physical photon becomes a
certain coherent superposition of the bare ordinary photon
and the bare mirror photon. Similarly, the physical mirror
photon is another, but {\it not orthogonal},
coherent superposition of the bare states. One consequence of this
is that purely ordinary stars cannot lose energy through mirror photon
emission.

We have discussed the propagation of mirror (ordinary) photons through an
ordinary (mirror) matter medium. This led us to consider qualitative
observational signatures of some hybrid ordinary/mirror binary
astrophysical systems. In particular, a mirror star
with an ordinary matter core can phenomenologically mimic a white dwarf,
and we have speculated that badly understood objects such as DC white dwarfs,
or the faint halo objects recently reported,
might actually be such hybrids.

\acknowledgments{This work was supported by the Australian Research
Council. RF is an Australian Research Fellow.}

\newpage
\section{Figure Captions}

\noindent
Figure 1: a) $A - A'$ mixing vertex. b) The interaction between the
ordinary electron and $A'$. c) The interaction between the
mirror electron and $A$. d) The interaction between the ordinary
electron and $A$.
\vskip 0.5cm
\noindent
Figure 2: Representation of the photon/mirror photon states.
Because $A$ and $A'$ are degenerate (both massless) any point on
the circle is a possible physical state in vacuum.
For propagation in ordinary matter the degeneracy is lifted,
with the states
$A_1$ (the photon) and $A_2$ (the orthogonal "sterile photon"
state) being the Hamiltonian eigenstates.
$A_2$ and $A_2'$ are the corresponding mirror states
for propagation in mirror matter. Note that $\tan \theta = \epsilon$.

\newpage
\epsfig{file=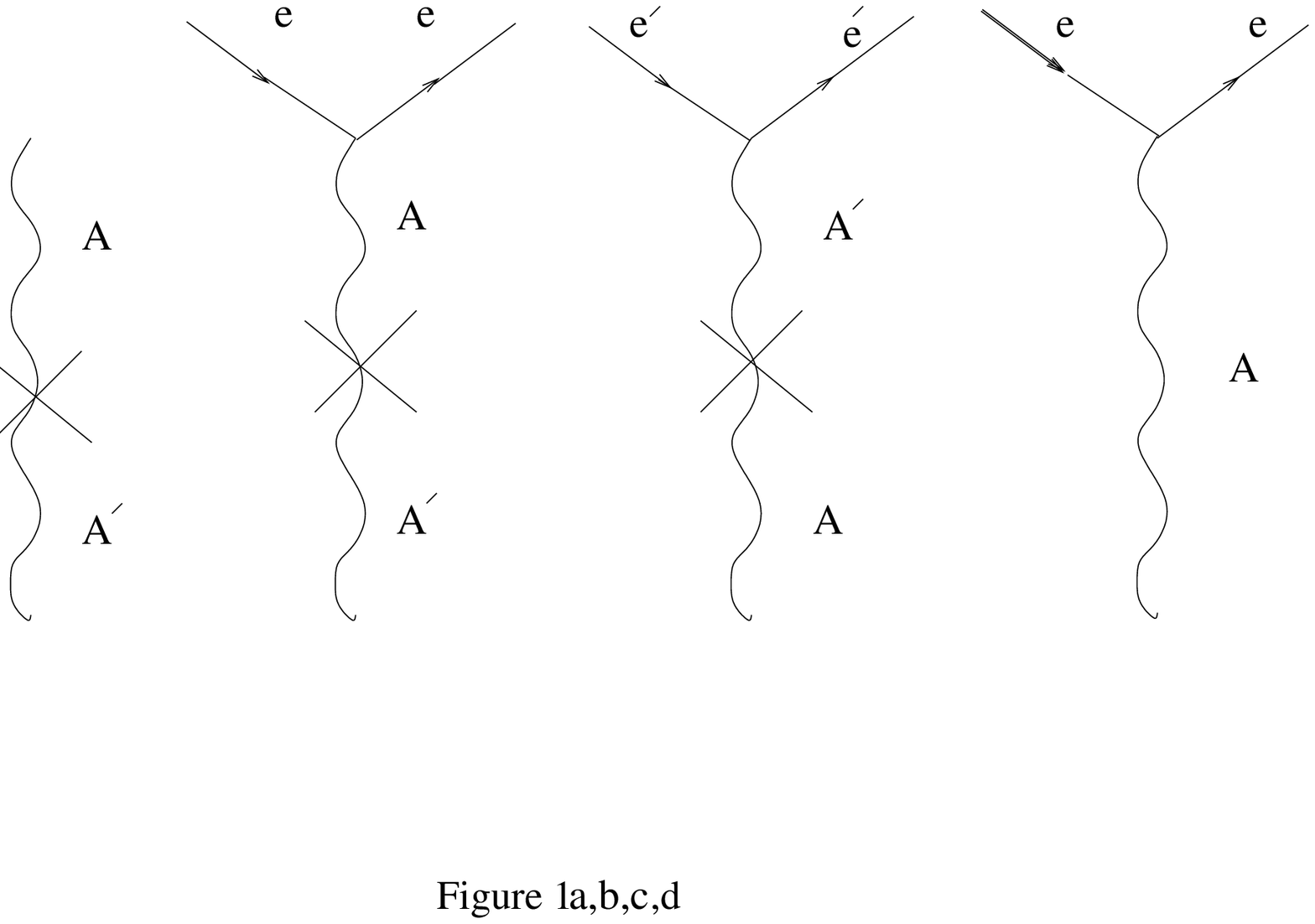, width=15cm}
\newpage
\epsfig{file=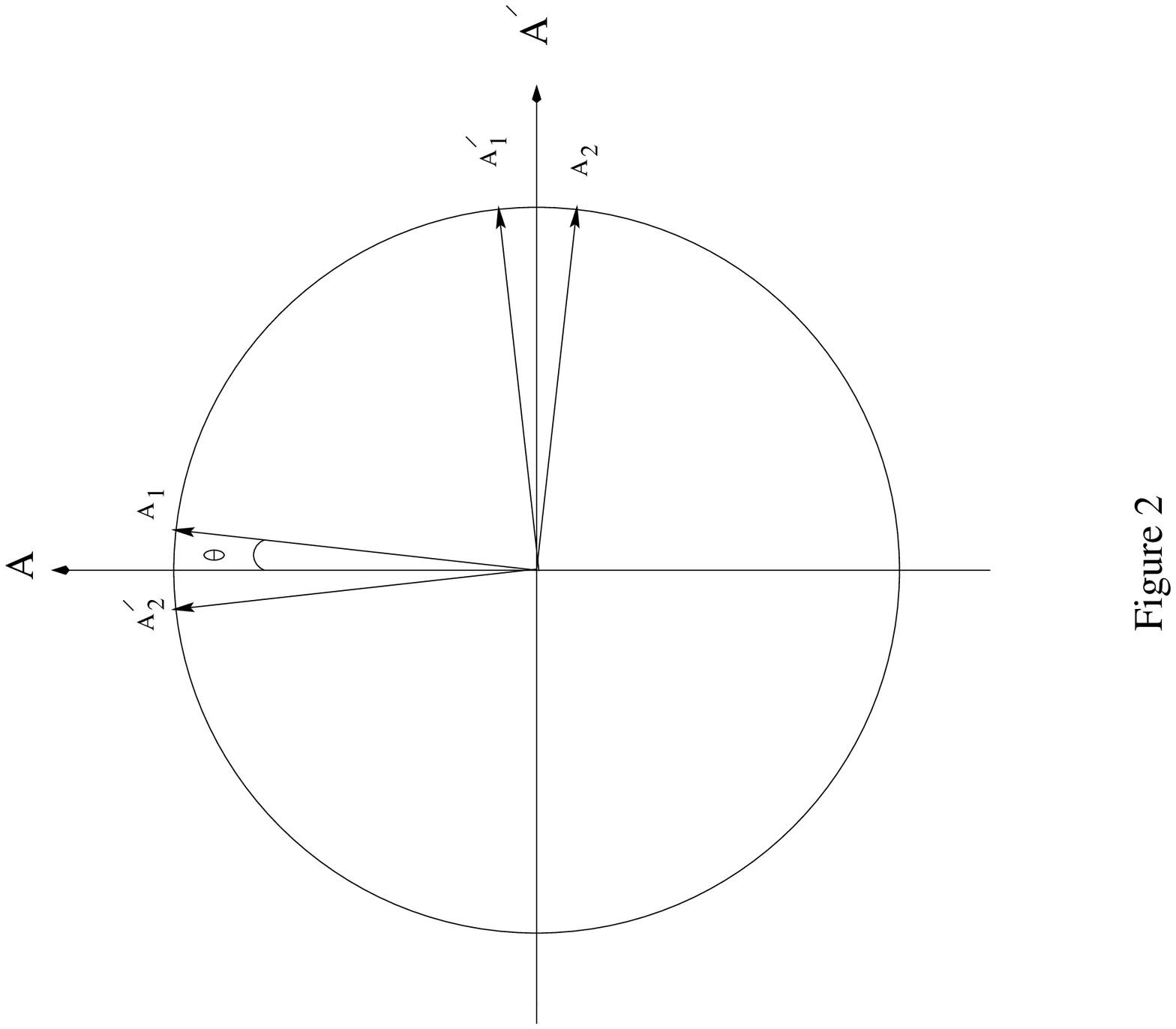, width=15cm}

\end{document}